\documentclass[preprintnumbers,amsmath,amssymb]{revtex4}
\usepackage{amsmath} 
\usepackage{graphicx} 
\usepackage{color}

\pdfoutput=1
\begin{document}

\title{Many-measurements or many-worlds? A dialogue.}

\author{Diederik Aerts}
\affiliation{Center Leo Apostel for Interdisciplinary Studies and Department of Mathematics, \\
Brussels Free University, Brussels, Belgium}\date{\today}
\email{diraerts@vub.ac.be} 
\author{Massimiliano Sassoli de Bianchi}
\affiliation{Laboratorio di Autoricerca di Base, 6914 Lugano, Switzerland}\date{\today}
\email{autoricerca@gmail.com}

\begin{abstract}

\noindent Many advocates of the Everettian interpretation consider that theirs is the only approach to take quantum mechanics really seriously, and that this approach allows to deduce a fantastic scenario for our reality, one that consists of an infinite number of parallel worlds that branch out continuously. In this article, written in dialogue form, we suggest that quantum mechanics can be taken even more seriously, if the \emph{many-worlds} view is replaced by a \emph{many-measurements} view. This allows not only to derive the Born rule, thus solving the measurement problem, but also to deduce a one-world \emph{non-spatial} reality, providing an even more fantastic scenario than that of the multiverse.


\end{abstract}

\maketitle

\section{Introduction}
\label{Introduction}

In this article we present a dialogue revolving around the interpretation of quantum mechanics and the origin of quantum probabilities. Two views of reality are confronted: a many-worlds view, in which all possible outcomes of a quantum measurement are always actualized, in the different parallel worlds, and a one-world view, in which a quantum measurement can only give rise to a single outcome. 

It remains indeterminate if this dialogue actually took place and, if so, where,  and who the two interlocutors exactly are. It is possible that they only exist, and will only exist, as imaginary characters in the abstract realm of human cognition. But it is also possible that in some of the worlds of a many-worlds reality, these characters not only have existed, or will exist, in a concrete objectual sense, but have also pronounced, or will pronounce, the exact words that are reported below. 

We leave it to the reader to decide, after having read the content of this conversation, which one of these two possibilities is more likely to be true, that is, if the following dialogue is the precise transliteration of an actual dialogue, which has happened, or will happen, with certainty, on a parallel Earth of a multiverse reality, or if it is just an imaginary conversation, which, although it could in theory have happened inside our reality, almost certainly has never been, and will never be brought into concrete existence as an actual -- fully incarnated -- conversation. 
\newline

{\scshape Tielo:} Hello \emph{Hudde}, I'm really happy that we have found some time to talk together again. What about picking up our exchange of ideas about the particular view of reality that we explored some time ago? I'm referring to the interpretation of quantum mechanics proposed in the 1950ies by the American physicists \emph{Hugh Everett III}~\cite{Everett1957}, and popularized by \emph{Bryce DeWitt}~\cite{DeWitt1973}, who asserts we all live in a \emph{multiverse}: a reality formed by \emph{multiple parallel worlds}. 

{\scshape Hudde:} With great pleasure \emph{Tielo}. I'm really glad to have this meeting with you, and yes, the Everettian interpretation is a perfect topic to discuss. If I remember correctly, what we both liked about it is that it takes the formalism of quantum mechanics very seriously, and that it can explain the quantum effects by allowing our physical reality to be much larger than just the four-dimensional spacetime.

{\scshape Tielo:} You remember well, and let me say that I was even more attracted by the multiverse view after having read a fascinating book by \emph{David Deutsch}~\cite{Deutsch1998}, in which he convincingly deduces, by analyzing the interferences produced by double-slit-like experiments, that the universe we see around us cannot constitute the whole of our reality, and that such reality necessarily contains vast numbers of parallel universes. So what I'm about to tell you may surprise you. Because you see, in my attempt to further my vision of such an ``enlarged reality,'' I made some unexpected discoveries, which have turned my perspective on the whole thing upside down. So much so, that I'm now no longer convinced that Everettians, or post-Everettians, really take quantum mechanics as seriously as they say they do. 

{\scshape Hudde:} This is surprising indeed. How did you come to that conclusion?

{\scshape Tielo:} Well, recently, and quite by chance I must say, I happened to read some engaging articles about the measurement problem, providing a surprising perspective on the whole issue; a perspective I subsequently summarized in the form of an unlikely question which has since been on my mind like a \emph{mantra}.

{\scshape Hudde:} I'm really curious about this question. What is it?

{\scshape Tielo:} The question is: Why was \emph{Laplace} not an Everettian?

{\scshape Hudde:} I beg your pardon?

{\scshape Tielo:} Yes: Why didn't Laplace consider the existence of a multiverse to explain the origin of its probabilities? What I mean to say is that probabilities are about \emph{possibilities}, and since possibilities can be \emph{chosen}, why didn't Laplace consider that each time this happens, the world would split into a number of parallel worlds, as many-worlders usually do? 

{\scshape Hudde:} This is a very strange question to ask. I'm sure you know that Laplace, and the other founding fathers of probability theory, which was beautifully axiomatized by \emph{Kolmogorov} in the 1930ies ~\cite{Kolmogorov1933}, were dealing only with \emph{classical probabilities}, whose nature is very different from \emph{quantum probabilities}. Well, not fundamentally different, but this we only know today, following the discovery of quantum mechanics, and following the observation that our reality is fundamentally quantum. But before the advent of quantum mechanics, classical probabilities, obeying Kolmogorovian axioms, certainly admitted a very different interpretation than quantum probabilities, which disobey these axioms and which emerged from our investigation of the microscopic particles.

{\scshape Tielo:} Yes, of course I know this, but tell me: Do you agree that, from a certain perspective, probabilities are about our \emph{lack of knowledge} regarding what can possibly happen in a certain portion of our reality, in a given context, whether such reality is a one-world or a many-worlds reality, and whether this lack of knowledge would be objective or subjective? I mean, do you agree with the idea that probabilities are \emph{epistemic} quantities? 

{\scshape Hudde:} As far as I'm concerned, I do not attach any specific \emph{ontic} character to probabilities in general.

{\scshape Tielo:} The same goes for me, and that's why I think it is not irrelevant to ask why Laplace, with his probability calculus, would have been in a fundamentally different situation than a modern quantum physicist. 

{\scshape Hudde:} But quantum physicists are confronted with the \emph{measurement problem}, whereas there was no such problem in the \emph{games of chance} investigated by the founders of probability theory. 

{\scshape Tielo:} Could it be that they simply did not perceive it as a problem, and therefore it wasn't identified as such?

{\scshape Hudde:} I'm not sure to understand what you mean by this. But before you go ahead, perhaps it would be useful to refresh our memories on the measurement problem. 

{\scshape Tielo:} Certainly.

{\scshape Hudde:} In short, quantum theory, in its standard Copenhagen formulation, affirms that there are two distinct processes describing the evolution of a quantum entity. One is a process of a continuous, deterministic kind, described by the \emph{Schr\"odinger equation}~\cite{Schrodinger1926}; the other one is a process of a discontinuous, indeterministic kind, governed by the \emph{projection postulate} and the associated \emph{Born rule}~\cite{Born1926,Neumann1955}. The first describes the unitary evolution of an isolated system, when it is not subjected to a measurement, while the second describes the effect of a measurement, in terms of an indeterministic \emph{collapse} (or \emph{reduction}) of the state vector. This introduces in our description of reality a dichotomy between measuring systems and measured entities, and a dualism in the processes of change of the state vector. As is well-known, when measuring systems are considered as systems that in turn can be measured by other measuring systems, and so on, this produces an infinite regress, which is a problem because the only way to break such infinite regress is to evoke the possibility of something inside reality which is not subjected to the laws of quantum mechanics, and which would act as the final measuring system. This is what von Neumann used to call the \emph{abstract ego}, and this is why many of those who adhere to the Copenhagen interpretation, perhaps without admitting it, consider that the \emph{consciousness} would play a special role in our reality. 

{\scshape Tielo:} Yes, and the Everettians, to solve the problem, postulate the existence of an infinite number of non-denumerably, increasingly divergent and non-communicating parallel quantum worlds, in which all possible quantum outcomes would be always actualized.

{\scshape Hudde:} To say the truth, they don't postulate the parallel worlds, they just deduce their existence from the theory. Their point is that all we need to know about a physical system, at a fundamental level, is contained in its state vector, and its evolution through the purely deterministic Schr\"odinger equation. We should therefore abandon the idea that there would be a different physical process associated with the collapse of the state vector. This is a choice of great simplicity: only one fundamental process of change would exist, namely the unitary and deterministic one described by the Schr\"odinger equation, which would also be applicable to the measurement processes. When a measurement system and a measured entity interact, they become entangled; the entanglement is the expression of a superposition of different states, quickly evolving, through the process called \emph{decoherence}, into a superposition of \emph{non-interfering alternatives}; these alternatives, or outcomes, all describe actual events, coexisting in different worlds, but considering that each version of the experimenter, in each one of these parallel worlds, is only aware of one of these outcomes, s/he will be erroneously induced in the error of believing that s/he lives in a single world. But this is just an illusion, resulting from the fact that her/his mind -- which also results from the activity of a physical system -- is split in a measurement process too. In other words, the laws of quantum mechanics would impose a sort of ``tunnel vision,'' creating the illusion of a single world, but these same laws also tell us that we have multiple simultaneous experiences, associated with different outcomes, all taking place in different parallels universes. 

{\scshape Tielo:} Thanks for this concise summary. Let me add that in the multiverse view, the statistical content of quantum mechanics, as expressed by the Born rule and confirmed by our laboratory experiments, would just be a consequence of the fact that we lack knowledge about which world of the multiverse we end up each time, following the ``branching process'' produced by a measurement interaction. 

{\scshape Hudde:} Precisely. No real collapse would take place. The multiverse would be described by a single state vector that Everett called the \emph{universal wavefunction}, evolving according to the Schr\"odinger equation, or to some \emph{linear} generalization of it. 

{\scshape Tielo:} The problem I have with the ``branching mechanism,'' describing the separation of the multiple worlds when a measured entity and a measuring system interact -- now that I have been reading these new articles that I will tell you about in much more detail later -- is that it more and more looks to me now like a disguised way of speaking of the projection postulate. Also, I find it disturbing that, to take quantum mechanics seriously, we have to eliminate from it something to which it owes much of its richness and subtlety, namely the projection postulate. 
I now understand how dangerous it is to solve a problem by eliminating the carrier of the problem: the operation may look successful at first, but what if the patient -- the theory -- has died? I agree of course that, mathematically speaking, and also from the viewpoint of an axiomatic operational approach to quantum mechanics, the Schr\"odinger equation is the simplest of all dynamics, only inducing an automorphism of the state space~\cite{Piron1976}. But why should we consider this simple dynamics to be the most general? Similarly, on what grounds should we consider the Hilbert space to be the most general structure for the set of states of a physical entity? In fact, what I have learned now is that the very existence of \emph{superselection rules}~\cite{Streater1964}, limiting the validity of the superposition principle, indicates that more general structures are likely to be at play within our reality. And that it is less naive to think that both the reversible evolution, governed by the Schr\"odinger equation, and the irreversible reduction of the state vector, are only idealized approximations of more complex processes of change. This seems very natural to me, as we undoubtedly live in a very complex reality, and different processes of change should be expected to unfold, depending on the context in which the different physical entities are immersed. No doubt some of these processes can be explained in terms of more fundamental ones, but can we really declare from the outset that the deterministic Schr\"odinger evolution would be more fundamental than the indeterministic von Neumann collapse? Can we really consider the former would be objective, and the latter, only subjective? Shouldn't we be more prudent before taking such a radical and dramatic stance? 

{\scshape Hudde:} I understand your caution, and I realize you have been reading new interesting approaches that have led you to reflect deeply about all this, following our past exploration of the many-worlds interpretation. But don't you think that we should be really daring, and take literally what the wave function tells us about reality, and the different correlations that can arise among its subsystems, at all possible levels? The many-worlds picture that can emerge from this is certainly shocking, but this is how things appear to be, at a fundamental level, although of course all this remains hidden to us, as we cannot directly feel the worlds splitting, in the same way that we cannot feel the rotation of planet Earth. 

{\scshape Tielo:} I can assure you that in thinking through the implications of these new articles that I read, it is not at all the idea of an enlarged reality that shocks me. On the contrary, I'm quite convinced that~\cite{Deutsch1998} ``reality is a much bigger thing than it seems, and most of it is invisible,'' so that ``the objects and events that we and our instruments can directly observe are the merest tip of the iceberg.'' But following these new readings and reflections, what I have now some difficulty to accept is how easily one omits certain parts of the theory and still pretends to take it very seriously. To really do so, shouldn't we accept all its postulates, although at first sight they may look problematic, and then investigate, without classical prejudices, their consequences for our view of reality? I still share with the Everettians, and with you, the desire to move beyond the antirealism subtended by the orthodox Copenhagen interpretation, and take very seriously what quantum theory tells us about reality. But as I said, I don't think any more today that the many-worlders take quantum theory \emph{seriously enough}! 

{\scshape Hudde:} Can you explain better?

{\scshape Tielo:} What I mean to say is that by eliminating the state vector collapse from the theory, one also eliminates the possibility of interpreting the \emph{superposition states} as \emph{genuine new elements of our reality}, reducing them to mere mathematical representations of collections of collapsed vector states, describing classical-like entities in distinct parallel worlds. 

{\scshape Hudde:} I see, but as far as I understand, to the Everettians, a so-called superposition state is in fact a genuine new element of reality, describing a \emph{superposition of worlds}. 

{\scshape Tielo:} Yes, I know this, but all these superposed worlds are just classic worlds, in which the \emph{potentiality} expressed by a superposition state is totally absent, as it cannot be associated with the actual state of any physical entity in any single world of the vast multiverse. That's why I affirm that in the many-worlds interpretation a superposition of states is not allowed to be a genuine new element of reality; I mean, a genuine new element of reality of a one-world reality. 

{\scshape Hudde:} I'm not sure what element of reality a superposition of states, in a one-world reality, would correspond with. Can you give me a simple example?

{\scshape Tielo:} Sure, but to avoid any misunderstanding, let me emphasize that when I say that we should take the projection postulate more seriously, I don't want to rule out the possibility that our reality could be governed by strictly deterministic processes. It is very possible, and I think also reasonable, to believe that indeterminism only arises as a consequence of our lack of knowledge, or control, about what really goes on, ``behind the scenes,'' when two physical systems are brought together and start interacting. 

{\scshape Hudde:} On that we totally agree.

{\scshape Tielo:} Great. Let me try then to explain what a superposition state is, according to my current understanding, which as I said is the result of my recent readings, of which I shall tell you more later. But for this I will have to go back to my curious question: Why wasn't Laplace, in his description of a game of chance, concerned by the unsupportable dichotomy of having to deal with both deterministic and indeterministic processes? 

{\scshape Hudde:} He wasn't because there isn't a state vector collapse in a gambling situation, nor a Born rule describing the probabilities of the different outcomes. Tossing a die on a table is not a measurement process. 

{\scshape Tielo:} But wasn't it Laplace who considered a principle to assign and calculate probabilities, known as the \emph{principle of insufficient reason}, or \emph{principle of indifference}?

{\scshape Hudde:} Yes, the principle was championed by Laplace, and if I remember well, also by \emph{Pascal}, \emph{Bernoulli}, and many others. It assigns probabilities in the absence of any evidence. Laplace considered it to be so obvious that he did not even bother to give it a name. 

{\scshape Tielo:} Unlike Laplace, let us do it, and call it, for the purpose of our discussion, the \emph{Laplace rule}. In his own words, his rule says that~\cite{Laplace1814}: ``The theory of chance consists in reducing all the events of the same kind to a certain number of cases equally possible [...] and in determining the number of cases favorable to the event whose probability is sought. The ratio of this number to that of all the cases possible is the measure of this probability, which is thus simply a fraction whose numerator is the number of favorable cases and whose denominator is the number of all the cases possible.'' Now, from my perspective, this Laplace rule, if correctly understood, can be considered as a limit version of the Born rule. I will tell you in a minute what I mean by that. For the moment, let me just draw your attention to the fact that, similarly to the Born rule, the Laplace rule is a \emph{rule of correspondence}, allowing to bring the ``theory of chance'' in contact with the experiments, which are the ``games of chance,'' with their unpredictable outcomes. I'm perfectly aware that Laplace wasn't describing the same layer of our reality as modern quantum physicists, but he was certainly also dealing with both deterministic and indeterministic processes, reversible and irreversible, the former being described by the Newton laws and those voluntary actions of human beings which were under their full control, and the latter being precisely the object of his theory of chance. Considering all this, I think it is very pertinent to ask why Laplace, as well as his predecessors and successors, didn't try to eliminate all indeterminism from their theory of chance, and just describe the possible outcomes of the different games in terms of multiple branching worlds. 

{\scshape Hudde:} I'm not sure if I follow the logic of your question. Is it really a pertinent one? How can you compare the experimental situation Laplace was confronted with, with that of modern quantum physicists? Also, Laplace wasn't studying the states and properties of physical systems, as physicists usually do. Furthermore, when you toss a coin, a die, or draw a ball from an urn, all aspects of the process are under your eyes: there are no mysteries, no cognitive problems, thus no need at this level to postulate multiple universes. 

{\scshape Tielo:} Are you saying that the many-worlds interpretation doesn't apply to macroscopic objects, like the sadly famous Schr\"odinger's cat, but only to microscopic ones?

{\scshape Hudde:} Of course not: if correct, it has to work at all levels of our reality. 

{\scshape Tielo:} So it should apply also to the different outcomes of a ``rolling experiment,'' in a typical dice game. 

{\scshape Hudde:} I think so, yes, because also when you roll a die, worlds will split, but this follows from the laws of quantum mechanics, not from the laws of classical mechanics. As you know, Everettians generally believe that the laws of \emph{classical} mechanics hold at a fundamental level, but because of the superposition of worlds, and the hidden interactions that these superpositions produce, the effective behavior of microscopic entities appears to be non-classical, that is, quantum, but this is just the result of this collective deterministic evolution of the mutually interacting and constantly splitting worlds. Since macroscopic entities are made of microscopic ones, the same holds true for them, but because of the phenomenon of decoherence, they don't show (in standard conditions) interference or non-locality phenomena, and they \emph{seem} to behave purely classically. But apart from this suppression of the quantum effects at the macroscopic level, the reason why we wouldn't see superposition states is of course the same for microscopic and macroscopic systems: it is because worlds split. 

{\scshape Tielo:} Yes, classical probabilists like Laplace only considered in their games of chance ordinary macroscopic objects, such as coins, dice, urns, etc., and therefore they couldn't observe interference effects and other quantum effects. This means they couldn't deduce the existence of superpositions of states, so that they couldn't infer the existence of the different branches of the multiverse, with all the possible outcomes always occurring in each of them. In addition to that, as you said before, they weren't measuring any physical property. 

{\scshape Hudde:} Precisely so.

{\scshape Tielo:} Of course, I do agree that, when we roll a die, everything, well, almost everything, is happening under our eyes. And this is precisely the reason why we had better observe this process not only with open eyes, but also with an open mind. For instance, it would be interesting to ask what kind of physics Laplace could have discovered if he had described the rolling of a die by using the concepts usually employed by physicists, like states, properties and measurements. 

{\scshape Hudde:} But rolling a die is not a measurement. 

{\scshape Tielo:} Usually it is not considered as such, I agree. But suppose it was, what kind of measurement would it be? 

{\scshape Hudde:} Undoubtedly a strange one. 

{\scshape Tielo:} In the same way a quantum measurement is a strange measurement, when compared to a classical one?

{\scshape Hudde:} Is it a pun?

{\scshape Tielo:} No, I'm just observing that if we consider the rolling of a die a process of measurement, then, according to our \emph{classical prejudice} of what a measurement should be, we are confronted with a strange situation, very similar to that in which the pioneers of quantum mechanics found themselves when they began to observe the microscopic entities.

{\scshape Hudde:} And what would be this classical prejudice?

{\scshape Tielo:} That a measurement process should always be a pure \emph{discovery} process, that is, a process through which we would take knowledge only of what already exists, in the actual sense of the word, prior to its execution. 

{\scshape Hudde:} I would say that this is not a prejudice, but the definition of a measurement. 

{\scshape Tielo:} Well, we may need to check carefully if our definitions are compatible with the structure of our reality, and the way we experience it. I'm not saying that measurements of a purely discovery kind would not be \emph{bona fide} measurements. What I'm saying is that they correspond to a very special kind of measurements, as measurement processes generally involve not only a \emph{discovery aspect} but also a \emph{creation aspect}. This is so because measurements can generally also change the state of the entity under consideration. 

{\scshape Hudde:} I certainly agree that a measurement, being the result of an interaction, can in principle affect the measured system, but only in the sense of producing a disturbance. The typical example of checking the pressure of a car tire comes to my mind: during the process it is difficult not to let out some of the air, so that the measured pressure may not correspond to the tire's pressure before the measurement. But it is certainly always possible, in principle, to reduce these disturbance effects by improving the observational techniques. So, when addressing the measurement problem at a foundational level, this is not something we should worry about.

{\scshape Tielo:} Yes, but the creation aspect which I think is inherent in the quantum measurement processes cannot be reduced to a mere disturbance effect: it would be an aspect of a more fundamental nature, which cannot be eliminated by a technological improvement for being incorporated in the very protocols which operationally define the properties to be measured. What I am saying, and sorry now for really playing on words, is that if a typical quantum measurement looks so \emph{str-ange}, when viewed from the perspective of our classical prejudices, this could be because it contains aspects related not only to the \emph{str}-ucture of the measured entity, which is what in part can be discovered, but also to its ch-\emph{ange}, that is, how such structure dynamically reacts, when it is acted upon in a certain way during an experiment. In a nutshell, a quantum measurement would not be only about the being of an entity, but also about its becoming. It would not be only about the properties it actually possesses, but also about the properties it is able to acquire, through the observational process. 

{\scshape Hudde:} I understand your point, but should we really consider a process that involves such a level of intrinsic invasiveness a measurement? For me, a measurement only consists in observing the actual properties of a physical entity. 

{\scshape Tielo:} I think this is more a question of terminology. I'm ok if you prefer to call a quantum measurement a ``quantum action'' or whatever, to distinguish it from a pure discovery process. Personally, I think we can reasonably upgrade the concept of measurement, and more generally of observation, to also include aspects of creation. But this is just a personal terminological choice. What is important is to clarify whether or not a so-called quantum measurement involves such creation aspect, and if this is sufficient to explain its apparent strangeness. 

{\scshape Hudde:} I don't really have a problem in associating the concept of creation with a measurement, but I'm not sure if it is correct to do it the way you do. It seems to me that you are taking the term ``creation'' in too literal a sense, that is, in the sense of the creation of properties that the measured system didn't possess before the measurement process. I'm not sure if I can agree on that. If for instance I consider the Everettian view, the only thing a quantum measurement ``creates'' is the entanglement between the measuring system and the measured entity, which in turn is responsible for the splitting of the universes associated with the different terms of the obtained superposition of states. These universes, or worlds, would however not be created by the measurement: they would already be there, and simply become separated through the measurement interaction. This is because in the many-worlds interpretation there is no reduction of the state vector, so that there is no creation of properties in the sense that you mean either. 

{\scshape Tielo:} Yes, this is what the Everettian description tells us. But let me explain that a different interpretation is possible, one which takes quantum theory even more seriously than Everettians, in the sense that it takes very seriously not only the Schr{\"o}dinger equation, but also the projection postulate, with its irreversible effect of change of the state vector during the act of measurement. Interestingly, as I learned thanks to my recent readings, in the same way that, by dropping the projection postulate, Everettians have to enlarge their reality, transforming it into a multiverse, by doing the opposite, that is, by keeping the projection postulate, we also have to face a considerable enlargement of our reality, but still remaining within a one-world description. 

{\scshape Hudde:} Now you've made me really curious about this one-world enlarged reality. 

{\scshape Tielo:} Let me come back to Laplace and consider the act of tossing a die on a table. I would like to describe such a process as a measurement. So, what is the observable that is being measured? As is usual in a dice game, it is the number of dots on the die's \emph{upper-face}. If the die in question is a traditional six-faced die, this means that the possible outcomes of the measurement of the upper-face observable are the six numbers 1, 2, 3, 4, 5 and 6. 

{\scshape Hudde:} The first objection that comes to my mind is that an upper-face value is not an intrinsic property of the die. I mean, a die certainly possesses six faces, each one showing a different number of dots, but none of them is an upper-face. So, we cannot really observe the upper-face of a die.

{\scshape Tielo:} Strictly speaking you are correct. But you see, I'm free to attach to a die all the properties I want, provided I do so in a meaningful and consistent way. I have no problems admitting that I'm considering properties of a non-ordinary kind, which usually are not accounted for in the classical description of macroscopic objects. The reason why I'm doing this is that I want to show that, in testing these properties, we are in a situation very similar to that of a quantum physicists dealing with microscopic entities. So, yes, you are correct, an upper-face is not a property of a die of the same kind as the property of having a mass, volume, a certain number of faces, etc. It is in fact a \emph{relational property}~\cite{Sassoli2014}, which is only actualized when the die is placed in a specific relation with another physical entity: the game table. 

{\scshape Hudde:} Are you suggesting that quantum observables would be relational, like the upper-face observable of a die?

{\scshape Tielo:} I'm not necessarily affirming this, although I'm neither denying this possibility. I'm just observing that an ordinary object like a die can possess more properties than those usually considered, and that without knowing it, the founding fathers of classical probability theory were precisely measuring such unconventional properties, and consequently, always without knowing it, they were actually studying quantum, or rather, quantum-like, processes. 

{\scshape Hudde:} I'm not sure if I can really see how, but please go ahead.

{\scshape Tielo:} What I would like to do now is to provide an operational definition of the upper-face observable, that is, specify how it is observed. It is very simple. If the die is already on the table, the observation simply consists in reading its value. If, on the other hand, the die is not on the table, one has to throw it on the table and, as soon as it has stopped rolling, read the value of the side that is facing upward, which is the outcome of the measurement. According to this experimental protocol, a die on the table is a die in an \emph{eigenstate} of the upper-face observable, as it is clear that the measurement of the die's upper-face will not change its ``on-table state.'' On the other hand, a die that is in an ``off-table state,'' does not possess an upper face, as you rightly observed. For it, having a specific upper-face is only a \emph{potential property}, not an actual one, which can be actualized only through the measuring process, when the die is tossed on the table. And since all the six faces of the die are \emph{available} to become an upper-face, it is perfectly correct to describe the state of a die which is not on the table as a \emph{superposition} of the six different upper-face eigenstates. It is a \emph{superposition state} in relation to the measurement of the upper-face observable, because an upper-face for the die has not yet been \emph{created}. But more importantly, and more simply, it is a superposition state because the die is not lying on the table. 

{\scshape Hudde:} It sounds a little strange to use the concepts of ``eigenstate'' and ``superposition state'' in relation to the roll of a die.

{\scshape Tielo:} I agree, but despite this strangeness, let us see where this way of describing the game of tossing a die takes us. If Laplace had described it as a measurement process, he would have observed that, apart from the special situations in which the die is already on the table, he was not in a position to predict the outcome of a measurement, even though he perfectly knew the state of the die before the measurement. In other words, he could not have associated his lack of knowledge about the outcome to incomplete knowledge of the die's state. However, he would also have observed that a probabilistic prediction of the different outcomes was nevertheless possible, in accordance with his Laplace rule. Let me add to this picture that Laplace was perfectly aware that in addition to these indeterministic processes, associated with the measurements of the upper-face, there were also deterministic ones, like in the moments when he decided to take the die from the table and put it in another place, in a perfectly controlled and reversible movement. In other words, this fictional Laplace I'm talking about, probably living in one of the parallel worlds described by DeWitt, would have been in a situation which is conceptually and structurally very similar to that of a quantum physicist confronted with the measurement problem. That's why I think it is worth asking if there is any chance that the Laplace of this hypothetical parallel world could end up relying on a many-worlds interpretation to solve his \emph{ante litteram} version of the measurement problem. 

{\scshape Hudde:} I have now the strange feeling of being in a superposition state of a cognitive kind: on the one hand, I perfectly see why you are asking this question, and on the other hand, I really don't see why you are doing so. But I'm still convinced that despite the conceptual and structural similarities that you have highlighted, your ``parallel Laplace'' was not confronted with any problem.

{\scshape Tielo:} I perfectly agree with you on that, and let me tell you why: although a die which is not on the table is undoubtedly in a state of potentiality with respect to a measurement of its upper-face observable, that is, a state characterizable as a superposition of outcome-eigenstates, it is also perfectly clear that it is in a perfectly well-defined state, always belonging to the same one-world reality. This one-world reality, however, is bigger than that of the game table: dice can be on-table, and have a well-defined upper-face value, but they can also be off-table, still remaining in the same world, which obviously extends beyond the borders of the table. So, I hope you will agree, a Laplace adopting an Everettian interpretation would look to his colleagues as rather an eccentric individual, pretending to live within the narrow confines of the surface of a game table, and believing that if a die can behave in a strange \emph{non-local} way, as if not always belonging to the table's space, this would be because many interacting table-worlds exist in the vast multiverse. Tell me, would you adhere to such many-tables view, if you were in his place? 

{\scshape Hudde:} What you describe is very interesting, and of course, I would not defend a ``multi-table worldview,'' if I was in that Laplace's place, and this for the very reason you evoke: I can perfectly see with my eyes that the die can exist in a higher-dimensional space which is not that of the table, so I don't need to hypothesize any ``parallel tables.'' But the fact that I can see the die when it is off-table is precisely the reason why I'm not sure your metaphor is really pertinent. 

{\scshape Tielo:} Or, conversely, you could say that this is precisely its interest: everything is under our eyes, and this is tremendously helpful in guiding our intuition. By the way, not only does the die example allow us to see that a superposition -- off-table -- state should be considered as a genuine element of a one-world reality, it also shows what happens during a quantum measurement, and how quantum probabilities are able to emerge. 

{\scshape Hudde:} You see, I agree that in your example the off-table die corresponds to an element of reality, and that within the paradigm of your discussion it cannot be understood as a mere mathematical representation of a collection of collapsed eigenvectors, associated with different ``table-worlds.'' But the problem with your example is that it is too simplified: it is really only a metaphor, almost an allegory, too distant from the Hilbertian structure revealed by quantum theory. Just to illustrate why I believe this curious upper-face measurement cannot seriously represent a quantum measurement, consider the Laplace rule. The probabilities of the six different outcomes are all the same: they do not depend on the state of the die before the measurement. By contrast, the quantum mechanical Born rule does describe a statistics of outcomes which depends on the pre-measurement state of the measured entity. 

{\scshape Tielo:} I agree with you that the die example is a very simple model, and that it is actually too simple. But it is not so because I cannot describe a more articulated experimental situation: I just wanted to introduce a minimal amount of elements to didactically illustrate this ``one-world point of view.'' And I also agree when you say that the Laplace rule is quite different from the Born rule. Consider, however, that, to correctly describe the statistics of the upper-face measurements, we cannot apply the strict Laplace rule, but already a modified form of it. Indeed, the probability ${\cal P}(n)$ for the outcome $n$, with $n = 1,\dots,6$, is equal to ${1\over 6}$ only if the pre-measurement state is an off-table state, but it is equal to $1$ if the die is already on the table, showing the upper face $n$, and equal to $0$ if it is on the table showing an upper face $m$ different from $n$. So, you see, we already are in a situation where the probability is not totally insensitive to the die's state prior to the measurement. This is because the upper-face measurement I have defined is a so-called \emph{measurement of the first kind}~\cite{Neumann1955}, that is, a measurement such that if a second identical measurement is conducted, immediately after the first one, the same outcome is obtained, with certainty. But I agree, this is not representative of what usually happens in a quantum measurement. In fact, we should call the upper-face measurement that I have defined a \emph{solipsistic} measurement~\cite{AertsSassoli2014a,AertsSassoli2014b}, in the sense of a process maximizing the creation aspect. 

{\scshape Hudde:} Solipsistic? What does the observer's mind have to do with all this?

{\scshape Tielo:} Absolutely nothing, I'm just using the term ``solipsistic'' in a metaphoric sense, to indicate that the measurement reveals almost nothing about the measured entity, considering that all states which are not eigenstates are statistically equivalent, as they produce exactly the same statistics of outcomes. In other terms, you cannot reconstruct, not even in part, a pre-measurement state which is not an eigenstate, by analyzing the statistics of the outcomes, and this means that the discovery aspect of the measurement is reduced to a minimum. Of course, there is a reason why the classical games of chance only involved solipsistic measurements: because historically they were associated with the wagering of money, and one wanted to avoid that the players, by controlling the pre-measurement state, could increase the probability of obtaining a specific outcome state. 

{\scshape Hudde:} Yes, this is the reason why they are called ``games of chance,'' and not ``games of skill.'' 

{\scshape Tielo:} Precisely. No level of control is allowed by the measurement protocol. Of course, since we are not concerned with gambling, but physics, we are free to remove this constraint and explore measurements that offer a better balance between creation and discovery. As you certainly noticed, I have not specified how the die has to be thrown on the table. This is because a human being cannot normally control the trajectory of the die, when it is thrown, and the most infinitesimal fluctuation can radically change the outcome. This is why all the faces can become an upper-face, with an equal probability. But it is certainly possible and easy to imagine more sophisticated protocols, allowing to control certain aspects of the rolling process, and not others, for instance by imposing that the die can only rotate around a certain predetermined axis. This, as you can imagine, will cause the Laplace rule to no longer apply, as the choice of a specific rolling axis, given an initial orientation of the die, can preclude certain faces from becoming upper-faces. Different rolling axes will then be associated with different upper-face observables, and since rotations in general do not commute, these different observables will not in general commute. And as I am sure you know, as soon as we have incompatible observables, the classical formula of \emph{total probability} can be easily violated, because of the presence of \emph{interference-like terms}. But one can even go further, and consider \emph{coincidence measurements}, performing experiments with couples of entangled dice, that is, dice which are glued together, or connected in some other way, allowing for the \emph{creation of correlated upper-faces}, when they are rolled simultaneously. And this mechanism of \emph{creation of correlations} can be easily exploited to violate \emph{Bell's inequalities}. 

{\scshape Hudde:} Are you saying that we can create interference effects by simply rolling a die along different axis, and violate Bell's inequalities by simultaneously rolling two dice attached together?

{\scshape Tielo:} Yes, this is what I discovered, not without some surprise I must say. In two of the new articles that I read and studied, I was made to doubt the many-world interpretation as the one that really takes quantum mechanics seriously, because the author was able to show that also a die can be considered to be a quantum entity, merely by subjecting it to certain experiments~\cite{Sassoli2013c,Sassoli2014b}. This led me to think how many quantum phenomena this alternative Laplace could have potentially discovered. And since I was also reading that brilliant book I mentioned you in the beginning of our conversation~\cite{Deutsch1998}, which defends the Everettian multiverse view, I started asking this curious question of mine. 

{\scshape Hudde:} Why wasn't Laplace an Everettian?

{\scshape Tielo:} Exactly.

{\scshape Hudde:} I now understand a little better your cognitive trajectory. 

{\scshape Tielo:} You see, this game table metaphor is really emblematic of the dramatic difference between an enlarged one-world view, in which genuine superposition states can exist, outside of the ``table-space,'' and a many-worlds view were only ``table-spaces'' exist. You can use as many ``parallel tables'' as you want, but this will never enable you to capture the reality of an off-table state, which is a fundamentally different state. Also, in a multiverse only made of tables, creations cannot exist. This is because an act of creation requires a superposition state, that is, an off-table state, and there are no such states in the multiverse. The reason is that many-worlders have eliminated from the theory the projection postulate, which precisely describes processes containing not only a discovery aspect, but also a creation aspect. 

{\scshape Hudde:} If I understand correctly, in the game table metaphor the table would correspond to our Euclidean space, am I correct?

{\scshape Tielo:} Precisely. The table would be the place of residence of the classical spatial entities, the theatre in which, before the discovery of quantum physics, all physical entities were believed to belong. However, if we take the metaphor seriously, our reality is too big to be staged in a three-dimensional spatial theatre. The microscopic entities, for instance, for most of their time, when not integrated into macroscopic structures, are in an ``off-space'' state, that is, a state describing a \emph{non-spatial} condition, usually indicated by the term ``non-locality.''

{\scshape Hudde:} I can see the parallel you are drawing between the measurement of the die's upper-face, and the measurement of the position of, say, an isolated electron. Because of the well-known phenomenon of the spreading of the wave function, we know that the state of the electron rapidly evolves into a superposition of position states, and if I take seriously your idea that a superposition state of this kind, which of course can no longer be associated with a spatial state, is nevertheless an element of our reality, I mean, an element of a one-world reality, then I must conclude, as you do, that the bigger part of this reality would be non-spatial, and corresponds in your metaphor to all the places a die can occupy in addition to those on the game table. 

{\scshape Tielo:} You perfectly got my point. 

{\scshape Hudde:} What I like about this view is that it also acknowledges, like the Everettian view, that our reality encompasses much more than what our eyes can see. 

{\scshape Tielo:} Yes, and that's why I was so interested in having this discussion about ``enlarged realities'' with you, as I remembered that we were both quite fascinated by the reality enlargement offered by the Everettian interpretation. But when I discovered that objects as simple as dice can ``go quantum'' too, the Everettian view started to appear much less convincing to me. How about you? 

{\scshape Hudde:} I cannot speak for Everettians, as I'm not sure if I'm one of them. My understanding of these subtle matters is still evolving, and will probably continue to do so for a long time. However, if I try to put myself in the shoes of a convinced many-worlder, I think I would tell you that I cannot consider changing my worldview, not even provisionally, on the mere basis of a clever metaphor. The many-worlds interpretation is founded on a sophisticated mathematical language -- the Hilbertian one -- discovered in our attempt to understand the physical reality at its most fundamental level, that of the microscopic constituents. There is nothing \emph{ad hoc} about the Hilbertian formalism, and I would go as far as to say that the many-worlds picture is the only realistic ontology which can be attached to it. Quite the contrary, your way of describing the experiments with the dice seems to me totally \emph{ad hoc}, a pure ``exercise in style,''  difficult to take seriously. 

{\scshape Tielo:} As a convinced Everettian, you were quite convincing. 

{\scshape Hudde:} Well, I think the objection I have just made is a serious one. 

{\scshape Tielo:} Sure. But tell me then, still remaining in the shoes of a convinced Everettian, What would make such an ``exercise in style'' more serious in your eyes, and less \emph{ad hoc}?

{\scshape Hudde:} Well, to start with, could you give an example of a macroscopic entity, along the lines of your metaphoric die example, but capable of perfectly simulating the behavior of a quantum system? I mean by this a perfect \emph{isomorphism}. Can you exhibit an explicit model from which the Born rule, and not some \emph{ad hoc} modification of the Laplace rule, can be fully derived and explained, such that a superposition state can be viewed as an objective condition of the entity under consideration, within the same one-world? If you can do this, then I will certainly start paying more attention to your perspective. 

{\scshape Tielo:} This is precisely what I also asked myself when reading those articles about the quantumness of a die. And much to my amazement, I discovered that macroscopic ``quantum machines''  had indeed been worked out in other articles of the collection I was reading and that made me doubt about many-worlds. And these quantum machines are able to fully model a quantum measurement process. A paradigmatic example is the so-called \emph{spin-quantum machine}~\cite{Aerts1986,Aerts1998,Aerts1998b,Aerts1999,Sassoli2013b}, describing measurements perfectly isomorphic to those performed on the spin of a spin-${1\over 2}$ entity. And as for the Born rule, it can also be derived in a simple way within the model. 

{\scshape Hudde:} 
Well, now you have my full attention, I'm really keen to know how this unlikely spin-machine works.

{\scshape Tielo:} As you will see, it is very simple. The entity that is subjected to the measurements is a material point particle, localized on the surface of a three-dimensional hollow sphere, of unit radius, so that the different states of the particle correspond to the different places it can occupy on such surface. 

{\scshape Hudde:} Is this a representation of the \emph{Bloch-sphere}?

{\scshape Tielo:} Precisely, and as you know, each point on the Bloch-sphere, also called the \emph{Poincar\'e-sphere}, is in a one-to-one correspondence with the spin state of a spin-${1\over 2}$ entity.

{\scshape Hudde:} I usually just consider the points of the Bloch-sphere as an abstract representation of the complex vectors of a two-dimensional Hilbert space, so it is a little bizarre to consider them as the positions of a real material point particle, but please continue. 

\begin{figure}[!ht]
\centering
\includegraphics[scale =.8]{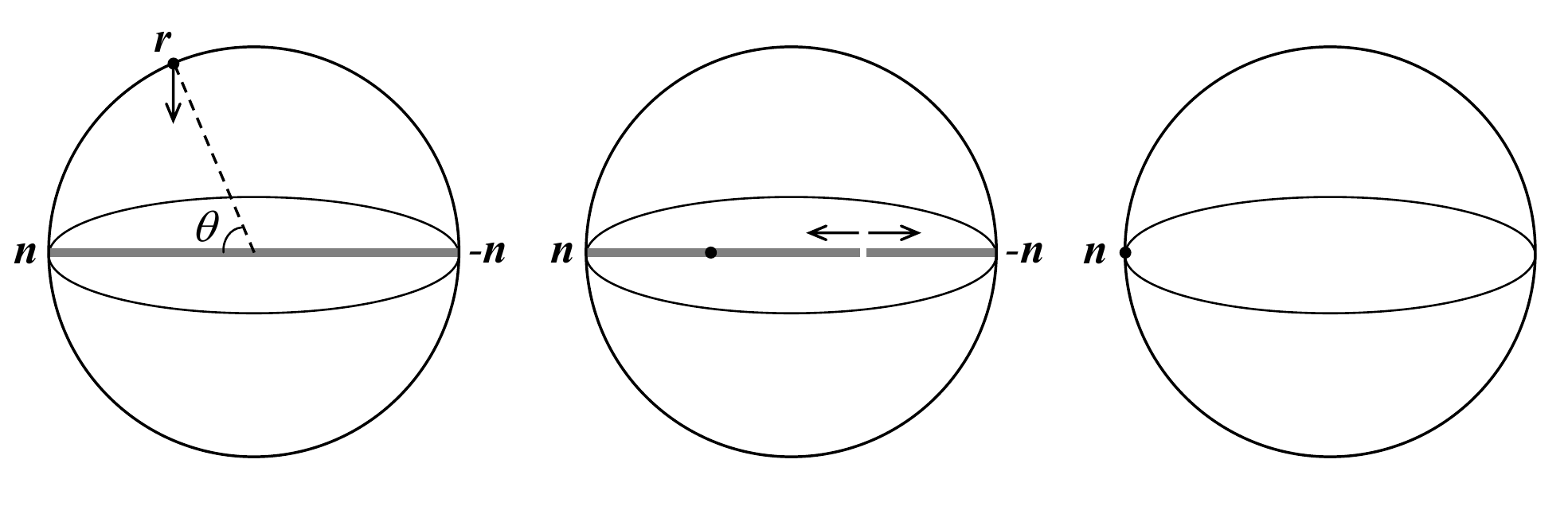}
\caption{The unfolding of a measurement process in the spin-quantum machine, with the point particle initially located in ${\bf{r}}$, orthogonally ``falling'' onto the sticky elastic band, which then breaks at some unpredictable point, and by contracting draws the particle to only one of the two possible outcomes, here ${\bf{n}}$.
\label{spinmachine}}
\end{figure}

{\scshape Tielo:} I should now tell you how measurements are \emph{operationally} defined on this material point particle. Each possible diameter of the sphere, identified by two opposite unit vectors $\bf{n}$ and $-\bf{n}$, defines a different observable $S_{\bf n}$. To measure $S_{\bf n}$, the procedure is the following: a uniform and sticky elastic band is stripped between $\bf{n}$ and $-\bf{n}$. Then, one lets the particle ``fall'' orthogonally from its pre-measurement location, specified by a vector $\bf{r}$, onto the elastic, firmly sticking to it. Next, one waits until the elastic breaks, at some unpredictable point, so that the particle, which is attached to one of the two broken pieces of it, will be pulled toward one of the opposite end points, $\bf{n}$ or $-\bf{n}$, corresponding to the outcome of the experiment, that is, the state acquired by the entity as a result of the measurement of $S_{\bf n}$. Using some elementary trigonometry, it is easy to calculate the transition probabilities for these two possible final states, and show that they exactly correspond to those obtained in a typical \emph{Stern-Gerlach} measurement. Indeed, the probability that the particle ends up in point $\bf{n}$ ($-\bf{n}$, respectively) is given by the length of the piece of elastic between the particle and the end-point $-\bf{n}$ (respectively, $\bf{n}$), divided by the total length of the elastic, which is twice the unit radius. Therefore, if $\theta$ is the angle between $\bf{n}$ and $\bf{r}$, we obtain that the probability for the outcome $\bf{n}$ (respectively, $-\bf{n}$) is given by: ${\cal P}_{\bf{r}}({\bf{n}}) =\frac{1}{2}(1+\cos\theta)=\cos^2\frac{\theta}{2}$ (respectively, ${\cal P}_{\bf{r}}(-{\bf{n}}) =\frac{1}{2}(1-\cos\theta)=\sin^2\frac{\theta}{2}$), which is exactly the quantum probability $ |\langle {\bf n}| {\bf r}\rangle|^2$ (respectively, $|\langle -{\bf n}| {\bf r}\rangle|^2$) obtained when measuring the spin observable $S_{\bf n}={\hbar\over 2}\left(|{\bf n}\rangle \langle{\bf n}| - |- \bf{n}\rangle \langle - \bf{n}|\right)$ on a spin-${1\over 2}$ entity prepared in the state $|{\bf r}\rangle$. And of course, if you repeat the measurement with the particle in one of the two extreme points of the elastic, the same outcome will be obtained, with certainty, so that the described process obeys von Neumann's ``first kind condition.'' I have made a little drawing of the process, so that you can better visualize it (see Fig.~\ref{spinmachine}). 

{\scshape Hudde:} Quite impressive, I must admit. The quantum probabilities, in this machine-model, result from the unpredictable breaking of the elastic band, am I correct?

{\scshape Tielo:} Exactly. To obtain the quantum probabilities, one has to use a \emph{uniform} elastic band, that is, a structure that has the same chance of breaking at any of its points. But of course, one is free to define even more general measurements, by also considering \emph{non-uniform} elastic bands~\cite{AertsSassoli2014a,AertsSassoli2014b,AertsSassoli2014c}. If for instance we use an elastic which can only break at its two end points, with equal probability, the two transitions will occur with probability $\frac{1}{2}$, irrespective of the pre-measurement state of the point particle. 

{\scshape Hudde:} Does this correspond to a solipsistic kind of measurement, as in your previous die example?

{\scshape Tielo:} Yes, it corresponds to a measurement where the creation aspect is maximized and, consequently, the discovery aspect minimized. 

{\scshape Hudde:} Can we also define measurements minimizing the creation aspect and maximizing the discovery one?

{\scshape Tielo:} Sure, they correspond to elastic bands which can only break at a predetermined point, so that it can be predicted if the particle, in a given initial state, will end up in $\bf{n}$ or $-\bf{n}$, if we exclude, of course, the exceptional circumstance of the particle falling exactly on the breaking point, as this would correspond to a situation of unstable equilibrium. Elastic bands that can break only at one point describe almost deterministic processes, called \emph{pure measurements}.

{\scshape Hudde:} Why pure?

{\scshape Tielo:} Because they are associated with a \emph{single measurement-interaction}. You see, each breakable point of the elastic corresponds to a different interaction between the elastic and the particle that is attached to it. This means that a uniformly breakable elastic incorporates a collection of different \emph{potential} pure measurements, each one associated with a different measurement-interaction. There is in fact a mathematical notion, called \emph{product experiment}~\cite{Aerts1982,Piron1976}, which I have learned provides a general definition of such a collection of potential pure measurements, but let us keep the discussion simple. Before orthogonally falling onto the elastic band, the point particle located in ${\bf r}$ is in a superposition state with respect to the measurement of $S_{\bf n}$, if ${\bf r}\neq \pm {\bf n}$. This superposition state corresponds to an element of reality of the ``spherical one-world'' of the particle, as we both can check with our own eyes, and is an expression of the fact that we cannot predict at which point the elastic will ultimately break. When the particle orthogonally falls onto the elastic, it experiences a sort of \emph{decoherence} process, and when it sticks onto the elastic its state is still the expression of a ``superposition of possibilities,'' for as long as the elastic remains unbroken. When it breaks, \emph{only one} among the potential measurement-interactions is actualized, and the point particle is drawn accordingly to \emph{only one} of the two possible end points. As was the case for the die experiment, there is obviously no need to hypothesize the existence of multiple interacting and branching worlds to explain the functioning of the spin-quantum machine, the existence of \emph{multiple measurement-interactions} being more than sufficient for this. In this way, not only can we keep our reality inside a one-world, but we can also enrich its fabric with the addition of the superposition states, describing a dimension of \emph{genuine potentiality}. This one-world reality does not contain every classical situation that one can imagine could have happened, or will happen: it is a reality where a clear distinction between ``what exists'' and ``what can possibly come into existence'' is allowed. In other terms, it is a reality in which our experiences, and experiments, can exhibit not only aspects of discovery, but also of creation, and in which breakable elastic bands are not confused with broken ones! Last but not least, it is a reality were the Born rule can be fully derived and explained as a uniform average over all possible measurement-interactions~\cite{Aerts1986,Coecke1995b,AertsSassoli2014c}. If I remember well, the derivation of the Born rule remains problematic in the many-worlds interpretation. 

{\scshape Hudde:} Some Everettians have claimed to have derived the Born rule by using arguments from decision theory~\cite{Deutsch1999, Wallace2003}, but as far as I know the validity of their derivation remains controversial, because of problems of circularity in their arguments~\cite{Barnum2000, Baker2007}. So, I'm really impressed by the fact that in this spin-quantum machine the Born rule can be derived in a very simple and objective way. For sure, if such a model could be taken seriously, as a description of a real quantum measurement, it would strongly challenge not only the many-worlds interpretation, suggesting its replacement by a \emph{many-measurement-interactions} interpretation, but also all other existing interpretations of quantum physics. 

{\scshape Tielo:} I also think that this derivation of the Born rule is quite something. Not only because, as far as I can judge, the many-worlds interpretation has been unable to convincingly derive it, but also because no interpretation of quantum mechanics seems to have ever succeeded in achieving this~\cite{Cassinello1996,Caves2005,Schlosshauer2005}, apart from, of course, this ``many-measurements'' interpretation I recently came across~\cite{Aerts1986,Coecke1995b,AertsSassoli2014c}. But you say that you can't take the spin model I have presented seriously. Why? All measurements have been defined in a very precise and consistent way, and the model is perfectly isomorphic to a spin-${1\over 2}$ quantum entity. It is not, of course, a spin-${1\over 2}$ quantum entity, but its set of states has exactly the same structure, and it possesses exactly the same observables. So, if a many-measurement-interactions interpretation allows to explain the origin of the quantum probabilities, as well as the reality of the superposition states in the case of the spin-quantum machine, why can't you consider that such explanation could also hold, \emph{mutatis mutandis}, for microscopic entities like electrons, protons, neutrons, etcetera? 

{\scshape Hudde:} Let me be clear, I think that the possibility of replacing the postulate of Born's rule by a genuine, non-subjective and non-circular derivation of it, would not only be an amazing goal to achieve within quantum theory in general, but would also indicate that the approach able to do so has good chances to provide the correct picture of what truly goes on during a quantum measurement. Now, the spin-quantum machine model is definitely intriguing, but I'm afraid it remains, like the die metaphor, an \emph{ad hoc} construction. 

{\scshape Tielo:} I'm confused. Before you told me that if I could show you a model that can do what the spin-quantum machine does, you would take the whole approach more seriously. What made you change your mind so rapidly?

{\scshape Hudde:} It's simple: while you were describing the model, I suddenly remembered that \emph{Simon Kochen} and \emph{Ernst Specker}, back in the 1960ies, were also able to construct an explicit macroscopic realization for the spin of a spin-${1\over 2}$ quantum entity~\cite{Kochen1967}. And I also remembered that they pointed out, on many occasions, that their model could only be constructed for a quantum entity with a Hilbert space of dimension not greater than two. In addition to that, I also know that the celebrated \emph{Gleason's theorem}~\cite{Gleason1957}, which affirms that the Born rule follows naturally from the complex Hilbert space structure, if vectors are assumed to provide a complete description of the states of a physical entity, is valid only for Hilbert spaces of dimension three or greater. As I am sure you are also aware, Gleason's theorem was instrumental in ruling out the existence of the hidden-variable explanations of quantum theory, but it didn't rule out such possibility for two-dimensional systems. Now, your spin-quantum machine is isomorphic to a two-dimensional system, and as far as I can judge it is a hidden-variable model. So, my educated guess is that this beautiful construction is just a two-dimensional anomaly, and that the many-measurement-interactions explanation cannot aspire to be of universal value, and supplant interpretations like the many-worlds one. Let me add that the only relevant hidden-variable theory which wasn't stopped by Gleason's theorem, and the other \emph{no-go theorems}~\cite{Neumann1932,Bell1966,Jauch1963,Kochen1967,Gudder1970}, is \emph{Bohmian mechanics}~\cite{Bohm1952b,Bohm1957}. This is because in Bohm's approach, as with the many-worlds interpretation, the collapse of the wave function is not considered to be a physical process, as it has no consequences for the trajectories of the point-particles described by the theory. But this is probably no coincidence, because Bohm's ``pilot wave'' can be interpreted as the wave function of the multiverse, guiding Bohm's single universe along its trajectory. In that respect, if I remember well, David Deutsch once said that~\cite{Deutsch1996} ``pilot-wave theories are parallel-universe theories in a state of chronic denial.''

{\scshape Tielo:} Now I understand better what causes your hesitation. All you say is of course correct, but you are confusing things a little. Let me explain. The spin-quantum machine model is indeed a hidden-variable model; however, the hidden variables are not associated with the state of the entity, but with the measurement interactions. This is why their introduction does not restore determinism, which is what physicists have historically tried to obtain when exploring hidden-variable models. In other words, the no-go theorems forbid the replacement of quantum mechanics by a more fundamental theory in which the entities would always be in eigenstates of all meaningful measurements, that is, by a fully deterministic theory in which the probabilities of having or not having a certain property can only take the values $0$ or $1$. This is not the case in the spin-machine model, which is a ``hidden-measurement'' model, and therefore cannot be stopped by the celebrated no-go theorems. It can, however, be stopped by the prejudice that makes many scientists think that an ideal measurement must necessarily be a fluctuation-free process, and that our physical reality must always be representable inside a spatial theatre. So, I hope I have reassured you that the no-go theorems do not apply to a many-measurements interpretation. 

{\scshape Hudde:} This is an important point, which I now see I have totally misunderstood. The spin-quantum machine is not about adding variables to the state of the entity, but about adding interactions to its measurements, and the no-go theorems  have no power to prevent this possibility.

{\scshape Tielo:} Exactly. But tell me, now that we have clarified this subtle aspect, are you better prepared to take this many-measurements interpretation as a serious contender for the description of the quantum reality?

{\scshape Hudde:} In a sense, yes, but you see, I'm still thinking that if it is possible to construct such a beautiful mechanical model for a spin-${1\over 2}$ entity, this is only because the two-dimensional Hilbert space can be represented in a real two-dimensional unit sphere. And even though, as you rightly observed, the no-go theorems cannot forbid one to build a many-measurements model beyond the two-dimensional case, say for a spin-$1$ entity, considering that the Bloch-sphere representation does not exist for Hilbert spaces of dimension three or greater, I remain quite convinced that this many-measurements mechanism is a sort of two-dimensional pathology. So, to answer your question, I think I would start taking really seriously this approach if it was possible to use it to generalize the derivation of the Born rule to an arbitrary number of dimensions. As you mentioned, no generally accepted derivation of the Born rule has been given so far, which of course does not mean that such derivation would be impossible, and if this idea of adding more measurement-interactions is able to do so, well, this I think would be a game changer.

{\scshape Tielo:} I agree, and the startling news is that the Born rule does in fact emerge in a very natural and general way from this many-measurement-interactions approach, which is something I was quite astonished to learn~\cite{AertsSassoli2014c}, also because the approach has remained almost unknown, and this despite the fact that the first encouraging results already appeared almost thirty years ago~\cite{Aerts1986}. 

{\scshape Hudde:} Then I think we should be very prudent. If it is so little debated among physicists, and has managed to do what all the other interpretations, apparently, have not been able to do, well, perhaps it is simply wrong. 

{\scshape Tielo:} This was indeed my first reaction. But then I started studying attentively the technical derivation, and to the best of my knowledge, there are no mistakes. 

{\scshape Hudde:} Then how do you explain this lack of awareness? 

{\scshape Tielo:} To be sincere, I don't know. My guess is that this could be because of a combination of factors. One of these could be historical, and related to the fact that this many-measurements interpretation emerged from the results of the \emph{Geneva school of quantum mechanics}, which used to be very active in operational and axiomatic research. It was not the only school investigating operational approaches to quantum mechanics, but I know that much of the research in this field took place in non-Anglo-Saxon countries. So, considering that today most of the foundational research happens within the very different perspective of the ``string-cosmological'' paradigm, of Anglo-Saxon origin, this may not have helped to bring more attention to these important ideas. Of course, psychological factors could also have played a role, like the difficulty in accepting that a measurement is not necessarily a discovery process, or that our reality cannot be staged inside too narrow a spatial theatre, be it three, four or eleven-dimensional, as assumed by some string theorists.

{\scshape Hudde:} But the many-worlders do accept the idea of a much vaster physical reality, which is typically infinite-dimensional. 

{\scshape Tielo:} You are correct, but remember that all these multiple branching worlds remain essentially ``table-worlds'' of a classical kind, and that there are no processes of creation in an Everettian multiverse. Anyhow, I think there is also an additional reason that may help explain this lack of awareness of the achievements of the many-measurements interpretation. By the way, I continue to call it this way, because in our discussion we have compared it with the many-worlds interpretation, but its historical name is \emph{hidden-measurement approach}, or \emph{hidden-measurement interpretation}. 

{\scshape Hudde:} So what would be this additional reason you are referring to?

{\scshape Tielo:} It is the same you also expressed, when you doubted that a hidden-variable model could be worked out beyond the two-outcome case, because of the no-go theorems. You see, as I have had the opportunity to learn, the proof that hidden-measurement models can be constructed for arbitrary quantum mechanical entities of finite or infinite dimensions has existed since decades~\cite{Aerts1986,Coecke1995b}. But it is also true that it was only for the two-dimensional case that a very natural and complete modelization of the measurement process could be given, using the Bloch-sphere representation. 

{\scshape Hudde:} You mean the spin-quantum machine you explained to me?

{\scshape Tielo:} Yes. I think that this lack of a natural generalization of the spin-quantum machine for higher-dimensional systems can in part explain the lack of interest in the hidden-measurement explanation. In the sense that, for those less familiar with the operational approaches to quantum mechanics, I can understand that the explanation remained difficult to evaluate.

{\scshape Hudde:} How is it then that in your case things went differently? It seems to me that you hold this approach in high esteem. 

{\scshape Tielo:} I do, and for the same reason we also held in esteem the many-worlds interpretation: I think it takes quantum mechanics very seriously. But in addition, and unlike the many-worlds interpretation and all other existing interpretations, it provides the only physically transparent and mathematically precise derivation of the Born rule that I know. This is not a minor aspect, if we consider that the Born rule is the core postulate of quantum mechanics. There is certainly also a more circumstantial reason why I find this approach attractive. I was lucky enough to discover a very recent work in which the authors do the very thing we have long been waiting for: provide this natural generalization of the spin-quantum machine~\cite{AertsSassoli2014c}. I can tell you that, while reading it, I had a moment of genuine epiphany: all of a sudden I could understand what a quantum measurement is. I was already quite amazed when the authors introduced the spin-quantum machine, corresponding to the two-outcome case, but I was even more amazed when they showed how the entire construction can be naturally generalized to an arbitrary number of outcomes. 

{\scshape Hudde:} Now I am getting really curious and interested. How can they do this, considering that the Bloch-sphere doesn't exist beyond the two-dimensional case?

{\scshape Tielo:} That's the point: it's not true, and this misconception is probably what has delayed the discovery of a natural higher-dimensional generalization. You see, the Bloch-sphere can be generalized to also represent states of an arbitrary number of dimensions. The  difference with the two-dimensional case is that not all points of the sphere will then be representative of states, that is, only a portion of the sphere will contain states, but such portion will still form a convex set. Before I explain to you, in very broad terms, how all this works, let me emphasize an aspect which apparently you didn't notice when I gave you a description of the spin-quantum machine. When the material point particle, initially at position ${\bf r}$ on the surface of the three-dimensional unit ball, is subjected to a measurement process, it enters the ball, reaches the elastic band and sticks to it. In the Bloch representation, the points inside the ball, as you probably know, are representative, not of vector-states, but of operator-states, also called \emph{density matrices}. 

{\scshape Hudde:} Yes, they are used in quantum mechanics to describe statistical mixtures of vector-states. 

{\scshape Tielo:} They can certainly be used in this way, but their interpretation as statistical mixtures of vector-states cannot be taken in too literal a sense, as is clear that a density matrix can have infinitely many representations as a convex linear combination of one-dimensional orthogonal projections, that is, as a mixture of vector-states. In other terms, the Hilbertian formalism already tells us that a density matrix cannot be consistently interpreted as the description of an \emph{actual} mixture, but only of a \emph{potential} mixture. In the spin-quantum machine, this is perfectly evident if we observe that the points inside the sphere, represented by non-unit vectors, can also be subjected to different measurements, and therefore cannot be intrinsically associated with a predetermined pair of outcomes. 

{\scshape Hudde:} I agree that since the internal points of the sphere also correspond to \emph{bona fide} states of the material point particle, and since these points are described in the Hilbert space language by density matrices, the latter should also be interpreted as pure states, at least in the ambit of this mechanical model. Personally, I have always considered the density matrices just as useful mathematical devices, and certainly not as descriptions of the reality of an entity. Are you suggesting that their interpretation as pure states should be general, and not limited to the description of the specific states of the point particle in the spin-machine? 

{\scshape Tielo:} If I take seriously the spin-quantum machine, as a modelization of what happens, in structural terms, during a quantum measurement, this seems to me an inescapable conclusion. And in principle it is also a testable conclusion, if one can find a way to probe the unfolding of a quantum collapse~\cite{AertsSassoli2014c}. Of course, there is no guarantee that this will be possible, as it could be that in the case of microscopic entities the process is not only non-spatial, but also non-temporal. Anyhow, let me now explain  how the spin-quantum machine naturally generalizes. I will limit my description to the simple case of a measurement having three outcomes, that is, of an observable having three eigenvalues. A typical example would be the spin of a spin-$1$ entity. I will also consider only the situation of a non-degenerate measurement. This is not because of any intent to hide difficulties: all I am going to tell you, beautifully and naturally generalizes to an arbitrary number of outcomes, also including the possibility of degenerate measurements. I just don't want to enter too much into technicalities, and keep our discussion mostly at the conceptual level, also because I'm sure that if you are intrigued by what I'm telling you, as I was, you will take the time to dig into the mathematics of the model~\cite{AertsSassoli2014a,AertsSassoli2014b,AertsSassoli2014c}.

{\scshape Hudde:} Sure, just explain to me in broad terms how the spin-machine model can be generalized, and I will trust you that you have checked the correctness of the maths behind the model.

{\scshape Tielo:} All right. What is important to observe is that the standard Bloch-sphere representation is based on the fact that the three traceless Pauli matrices, together with the identity operator ${\mathbb I}$, form a basis for the set of all linear operators on ${\mathbb C}^2$. Therefore, a density matrix can always be expanded on such a basis, and since it is self-adjoint, and of unit trace, it can generally be written in the form: $D({\bf r}) = {1\over 2}\left(\mathbb{I} + {\bf r}\cdot\mbox{\boldmath$\sigma$}\right)$, where \mbox{\boldmath$\sigma$} is a three-dimensional vector whose components are the Pauli matrices $\sigma_i$, $i=1,2,3$, and ${\bf r}$ is a real vector belonging to the three-dimensional unit ball. When ${\bf r}$ is of unit length, $D({\bf r})$ is a one-dimensional projection operator, whereas when its length is less than $1$, it is a density matrix. 

{\scshape Hudde:} Yes, this bijection between the operator states $D({\bf r})$, and the real vectors ${\bf r}$, belonging to the three-dimensional unit ball, is precisely the expression of this well-known homomorphism between $SU(2)$ and $SO(3)$. I wasn't aware, however, that it could be generalized to higher dimensions. 

{\scshape Tielo:} Not the homomorphism, of course, but you can still represent the quantum states inside a real unit ball and, consequently, the measurements as ``breakable elastic structures'' operating inside of it. Let me consider the example of ${\mathbb C}^3$. Also in this case, one can find traceless matrices which, together with the identity operator, form a basis of all linear operators acting on ${\mathbb C}^3$, so that one can still write a density matrix in the form: $D({\bf r}) = {1\over 3}\left(\mathbb{I} + \sqrt{3}\,{\bf r}\cdot\mbox{\boldmath$\lambda$}\right)$. In the same way as the three Pauli matrices are the generators of $SU(2)$, the eight components $\lambda_i$, $i=1,\dots,8$, of the vector \mbox{\boldmath$\lambda$}, are now the generators of $SU(3)$, and correspond to the so-called \emph{Gell-Mann} matrices. Since we have eight generators, the real vector ${\bf r}$ representative of the state of the entity is this time eight-dimensional. As I said before, an important difference with respect to the two-outcome case, in addition to this increase in the number of dimensions, is that the ball is not completely filled with states. This is so because not all its vectors can now be associated with a positive operator $D({\bf r})$.

{\scshape Hudde:} I see, to still have a Bloch-sphere representation for the states, beyond the two-dimensional case, the price to pay is that the spherical symmetry is lost, because of the absence of a homomorphism between $SU(3)$ and $SO(8)$. 

{\scshape Tielo:} Yes, this is how things are, but this lack of spherical symmetry is not per se a problem. In fact, what is important for the modelization of the measurement processes is that the vectors representative of the states form a closed convex set, which is always the case. Let me explain how it works. A non-degenerate observable is characterized by three distinct outcomes, associated with three distinct eigenstates. Let me denote ${\bf n}_1$, ${\bf n}_2$ and ${\bf n}_3$ the three unit vectors associated with these three orthogonal eigenstates. It is possible to demonstrate that inside the eight-dimensional ball they form an inscribed \emph{equilateral triangle}. This two-dimensional triangular structure represents the measurement context associated with the observable in question, in the same way as the one-dimensional elastic band represented the measurement context in the two-outcome situation. 

{\scshape Hudde:} Are you saying that you can associate a sort of elastic membrane to this equilateral triangle, and describe the measurement process as you did before with the elastic band? 

{\scshape Tielo:} Exactly. 

{\scshape Hudde:} That would be amazing.

{\scshape Tielo:} I agree with you. So, this time, instead of a one-dimensional uniform elastic band, we have a two-dimensional uniform elastic membrane. As before, during the measurement, the point particle, initially at position ${\bf r}$, orthogonally ``falls'' onto the membrane, and sticks to it. In this way, it defines three different triangular subregions on the membrane, delineated by the line segments connecting the particle's position with the three vertex points. You have to think of these line segments as ``tension lines'' altering the physics of the membrane, in the sense of making it less easy to break along these lines. Now, once the particle is on the membrane, firmly attached to it, at some moment the latter will break, at some unpredictable point belonging to one of these three subregions. The tearing then propagates inside that specific subregion, as a sort of disintegrative process, but not in the other two subregions, because of the tension lines. This causes the two anchor points of the disintegrating subregion to also tear away, thus producing the detachment of the membrane, which, being elastic, then contracts towards the only remaining anchor point, also drawing to that position the point particle attached to it, which then reaches its final state, corresponding to the outcome of the measurement. Could you visualize the process I'm describing?

{\scshape Hudde:} I think so yes, but a drawing would certainly be helpful. 

\begin{figure}[!ht]
\centering
\includegraphics[scale =.8]{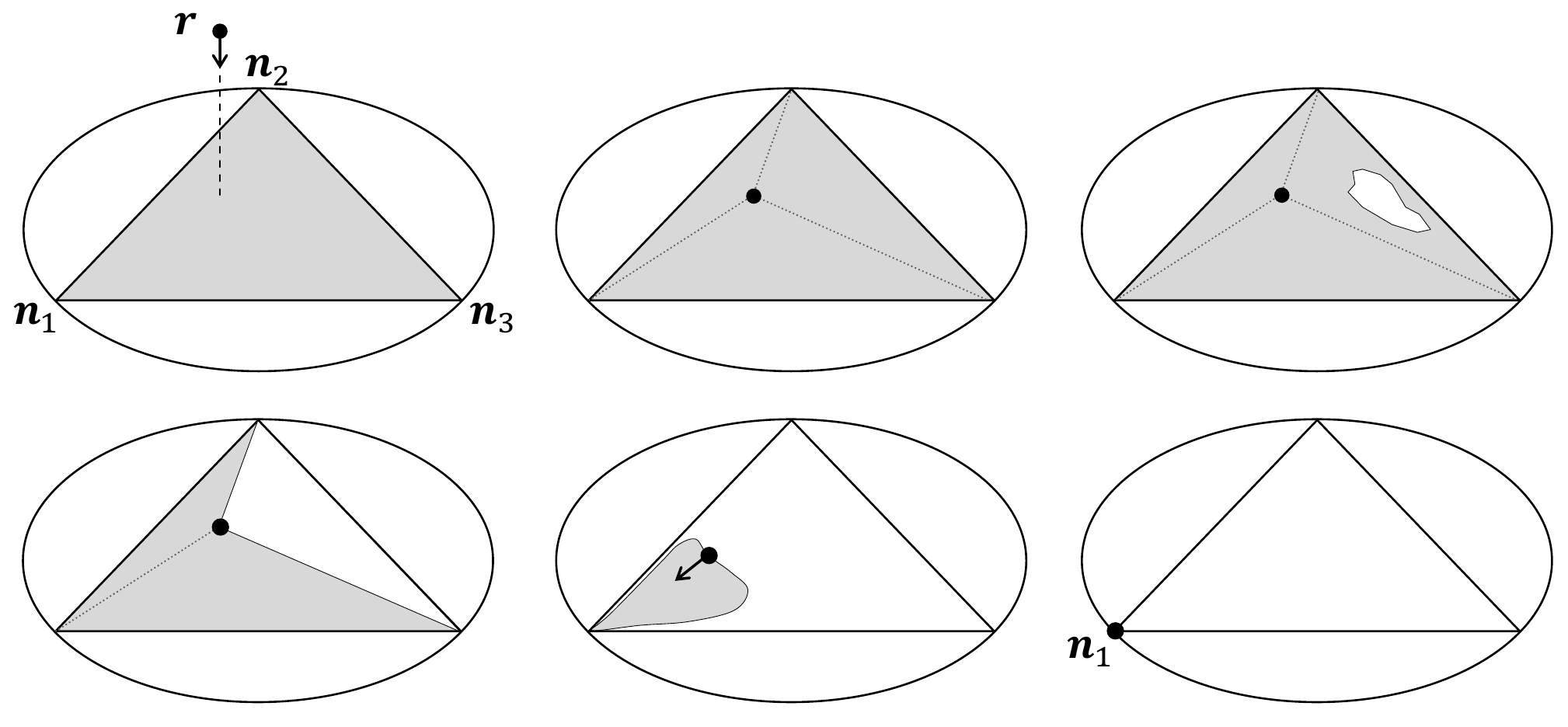}
\caption{The unfolding of a (non-degenerate) measurement process with three distinguishable outcomes: ${\bf n}_1$, ${\bf n}_2$ and ${\bf n}_3$. The point particle representative of the state, initially located in ${\bf{r}}$, orthogonally ``falls'' onto the triangular elastic membrane, thus defining three different subregions. The membrane then breaks in one of them, producing its disintegration and the subsequent detachment from its two anchor points, thus drawing the point particle to its final location, here ${\bf{n}}_1$.
\label{spinmachine3d}}
\end{figure}

{\scshape Tielo:} Sure, let me assume for this that the final outcome corresponds to the vertex vector ${\bf n}_1$. Of course, I cannot draw the full picture of an eight-dimensional ball, but I can certainly represent the two-dimensional circular section containing the triangular membrane. And by the way, while I'm drawing the picture  (see Fig.~\ref{spinmachine3d}), let me ask you: don't you find it amazing that it is possible to visualize in this way a quantum measurement, belying the widespread preconception of quantum processes being impossible to imagine, unlike classical processes? And this time, the process involves three possible outcomes, so that we are no longer in the context of a possible two-dimensional anomaly. 

{\scshape Hudde:} I agree that this is quite unexpected and seems to invalidate Feynman's famous quote that nobody understands quantum mechanics~\cite{Feynman1992}. But how does the Born rule come out from what you describe?

{\scshape Tielo:} In a very easy way. Each possible breaking point of the triangular membrane is associated with a potential pure measurement, which can bring the particle into a given eigenstate. These pure measurements are all deterministic, apart, of course, those associated with breaking points exactly located on the tension lines, as they correspond to situations of unstable equilibrium, where it remains indeterminate which subregion will disintegrate. But we don't have to worry about these exceptional measurement-interactions, as they are of zero measure in the determination of the outcome probabilities. To calculate them, we only have to observe that the membrane, by assumption, being a uniform structure, the probability of it breaking in any given subregion is simply given by the ratio between its area and the total area of the membrane. If you do this calculation properly, and here of course you will have to trust me, you will obtain exactly the Born rule. In other words, the uniform membrane measurements are perfectly isomorphic to the measurements of a non-degenerate observable in a three-dimensional complex Hilbert space.

{\scshape Hudde:} And you say that this ``multi-interaction'' scheme can also be generalized to an arbitrary number of outcomes, and that in addition to that you can also describe the situation of degenerate measurements? 

{\scshape Tielo:} Yes, this is what the authors of this article I was telling you about were able to do~\cite{AertsSassoli2014c}. For the general situation of an arbitrary number $N$ of outcomes, the Bloch sphere representation is $(N^2-1)$-dimensional, and the elastic structure describing a measurement becomes a $(N-1)$-dimensional \emph{simplex}, always inscribed in the sphere, or rather, in the \emph{hypersphere}. Degenerate measurements are a little more complex, in the sense that they correspond to situations where a certain number of subregions, precisely those associated with the degenerate eigenvalues, are fused together and form bigger composite (not anymore convex) subregions. When the initial breaking point takes place inside one of such composite subregions, the process draws the point particle not to a vertex point of the simplex, but to one of its sub-simplexes. The collapse remains perfectly compatible with the quantum mechanical \emph{L\"uders-von Neumann projection formula}, but to complete the process the particle will also have to re-emerge from the membrane, to deterministically reach its final position. In other words, in the general case, the measurement process is described by a tripartite process formed by (1) an initial \emph{decoherence-like} process, corresponding to the point particle deterministically reaching the ``on-membrane region of potentiality;'' (2) a subsequent indeterministic \emph{collapse-like} process, corresponding to the breaking of the membrane and the particle being drawn to some of its peripheral points; and (3) a possible final deterministic \emph{purification-like} process, bringing again the point particle to a unit distance from the center of the sphere~\cite{AertsSassoli2014c}. Of course, all I have said can be expressed in very precise terms, and certainly the entire explanation becomes much more convincing when one goes through all the mathematical aspects of the derivation. But for the purpose of our informal exchange, I think I have already given you enough information to ponder. 

{\scshape Hudde:} That's for sure. So, if all you have explained to me is correct, it follows that this hidden-measurement mechanism cannot be considered as a two-dimensional anomaly, but truly as a solution to the measurement problem. 

{\scshape Tielo:} Considering that, to the best of my knowledge, it is the only existing -- non-circular -- derivation of the Born rule, I think we can rightly say so. In fact, let me mention this en passant, it is even possible to relax the hypothesis that the membranes have to be uniform. If we take the average of all possible non-uniform membranes, it turns out that also in this case the Born rule can still be deduced~\cite{AertsSassoli2014b,AertsSassoli2014c}. But tell me now, do you think that it is always a good idea to eliminate from the start the projection postulate from the quantum formalism, as proposed by Everett? 

{\scshape Hudde:} Well, if the maths in support of all you have described is solid, I think I will have to consider this one-world hidden-measurement interpretation as a very serious rival of the many-worlds interpretation. So, in essence, if I understand correctly, if the projection postulate is maintained as a fundamental ingredient of quantum theory, the multiplicity of worlds should be replaced by a multiplicity of measurement-interactions, and by equipping a one-world reality with all these additional measurement-interactions, it automatically becomes a much larger reality, also including non-spatial entities. 

{\scshape Tielo:} Yes, and this you can see already in the three-outcome example, which requires an eight-dimensional unit ball to represent the reality of the entity in question, and that of its possible measurement contexts. No three-dimensional spin-quantum machine will ever be able to simulate the full working of a measurement situation beyond the two-outcome case. In that respect, you were right when you said that the two-dimensional situation was pathological. It was so, not in the sense that a hidden-measurement interpretation would not be viable beyond the two-dimensional case, but in the sense that the description of a quantum entity certainly requires to go beyond the ordinary three-dimensional spatial theatre. Well, this was in fact already the case for two-dimensional entities, if we observe that the rotation group $SO(3)$ is covered \emph{twice} by $SU(2)$, but let's not enter now into these more subtle questions, which are about understanding how the three-dimensional space that we perceive with our ordinary senses, like the sense of touch~\cite{Aerts2014d}, emerges from a genuinely higher-dimensional -- possibly infinite-dimensional -- non-spatial substratum. Speaking of multidimensionality, let me say that for me the fantasticalness of the Everettian worldview is largely overestimated, as a one-world view which takes seriously the projection postulate produces an even more fantastic reality. I'm saying this because, as I already emphasized in our conversation, all the worlds of the many-worlds interpretation are just classical worlds. Therefore, in the Everettian view one is somehow cloning the same classical structure \emph{ad infinitum}, and this of course does not really add anything new to reality, in structural terms. Of course, the different parallel worlds are not totally parallel, as each universe interacts with all the others, through interference phenomena. But nevertheless, the overall picture remains desperately classic. 

{\scshape Hudde:} I imagine that a convinced Everettian, more expert than me, would have many things to say in that respect. Personally speaking, although I'm a sympathizer of the many-worlds interpretation, I'm also very open to consider alternative approaches, and this hidden-measurement interpretation you have recently discovered is undoubtedly a serious contender, and something that I will certainly investigate more attentively. I must say that this idea of most of our reality being non-spatial is quite fascinating, and it is very possible that the multiverse might just be a clumsy way of trying to represent the nature of these non-spatial entities, which I hope you agree remain quite mysterious. 

{\scshape Tielo:} I do, but considering this non-spatial perspective, don't you think that it is the existence of a spatial ``window,'' in which we humans live with our macroscopic bodies, and through which we are able to detect non-spatial and higher-dimensional entities, that really constitutes the mystery to be explained? How does our three-dimensional space, or four-dimensional spacetime, emerge from such higher-dimensional, possibly infinite-dimensional, non-spatial realm?

{\scshape Hudde:} This is an interesting reversal of perspective. I really must thank you for this stimulating exchange of ideas. It would be nice if we could meet again, after I have meditated on all we have discussed. 

\begin{figure}[!ht]
\centering
\includegraphics[scale =.5]{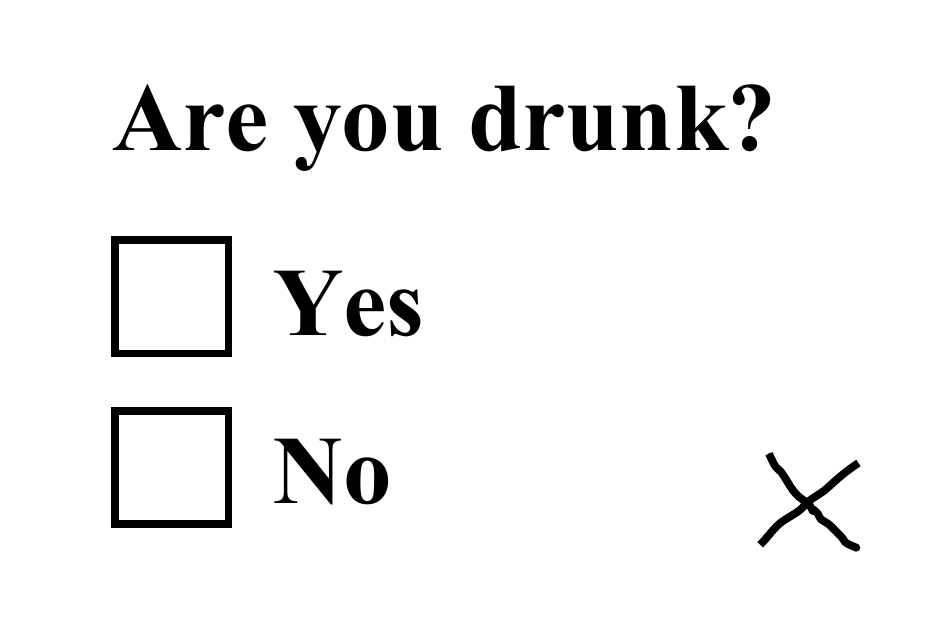}
\caption{A superposition state.
\label{areyoudrunk}}
\end{figure}

{\scshape Tielo:} Thanks to you, and yes, it is an excellent idea to meet again and confront the evolution of our views. But before we go, and to conclude our conversation on a funny note, let me show you a picture I received today, through my social network, as I suddenly realized how pertinent it is in relation to what we have discussed  (see Fig.~\ref{areyoudrunk}).  As you can see, the picture shows two boxes of a typical multiple-choice questionnaire. Only two answers are allowed -- ``yes'' and ``no'' -- and above the two boxes there is the question: ``Are you drunk?''

{\scshape Hudde:} A very funny picture.

{\scshape Tielo:} Yes, its humorous effect is produced by the presence of the cross in a ``place'' one would not expect. 

{\scshape Hudde:} Indeed, but I don't really see what this picture has to do with our discussion.

{\scshape Tielo:} A lot. It describes a \emph{yes/no-experiment}, which in quantum mechanics is associated with an orthogonal projection operator: an observable having only two eigenvalues, $1$ and $0$. In the present case, the observation is about the property of ``being drunk,'' and the eigenvalue $1$ corresponds to a cross placed in the ``yes box,'' while the eigenvalue $0$ corresponds to a cross placed in the ``no box.'' But let me ask you, why do you find the picture funny? 

{\scshape Hudde:} Obviously, because it shows that the guy who answered the question was drunk. 

{\scshape Tielo:} Well, this is a strange statement, considering that the property of ``being drunk'' corresponds to a cross drawn in the ``yes box.''

{\scshape Hudde:} Now you are confusing me.

{\scshape Tielo:} I will explain, but let me first emphasize that according to the ``standard rules of compilation of a questionnaire,'' only one of the two boxes can be checked with a cross. This reflects the fact that one cannot be drunk and sober at the same time, in the same way that the infamous Schr\"odinger's cat, as far as we know, cannot be dead and alive at the same time. This is the crucial difference between a one-world reality and a many-worlds reality. In the first, only one check is possible, whereas in the second both boxes are always checked. 

{\scshape Hudde:} But not both in the same world. 

{\scshape Tielo:} Yes, to avoid the non-sense of two mutually excluding possibilities being simultaneously actualized, a many-worlder will consider that instead of one pair of boxes there are two pairs of boxes, so that there is a drunk person in one world and a sober person in another. 

{\scshape Hudde:} Exactly, instead of one questionnaire there are two of them. 

{\scshape Tielo:} Well, the problem is that in the multiverse, questionnaires do not exist. 

{\scshape Hudde:} I don't think I understand.

{\scshape Tielo:} Only small boxes exist in an Everettian's multiverse, which, by the way, with their four sides,  perfectly symbolize the events of a four-dimensional spacetime. The problem that is posed by this ``misplaced'' cross is that it is located outside space, or spacetime. This out-of-box cross cannot exist in an Everettian multiverse. But if we take this picture seriously, we cannot deny its existence, as it has been drawn precisely there. 

{\scshape Hudde:} This reminds me of the off-table states in your die experiment. So, how should I understand this out-of-box cross? As a superposition state?

{\scshape Tielo:} Precisely. The off-box cross reminds us that humans can be in states that cannot be described by simply saying ``I'm drunk'' or ``I'm sober.'' 

{\scshape Hudde:} If it is a superposition state, then it should be possible to bring it into one of the two boxes, by means of a measurement. 

{\scshape Tielo:} Do you realize that by saying this you are no longer thinking as an Everettian? 

{\scshape Hudde:} Apparently, I have been contaminated by the many-measurements view. 

{\scshape Tielo:} If the cross represents a superposition state, describing the state of the person before the yes/no-experiment, then of course you can imagine that by collapsing a ``mental membrane,'' it can be moved from that off-box position to an on-box one. But I think there is more in the metaphor expressed by this picture. The reason it is so funny is that it depicts a sort of paradox, which reveals an interesting possibility, namely that not all states of our non-spatial reality are necessarily spatializable, in the course of a measurement. As we observed, the cross cannot be associated with the state of somebody who is drunk, but what about somebody who is ``very drunk''? When we ask them if they are  drunk, using a questionnaire, we will not be able to obtain any meaningful outcome. But this doesn't mean they are not in a well-defined and objective state. 

{\scshape Hudde:} Are you saying that not all superposition states would be collapsible into a spatial state?

{\scshape Tielo:} Think about \emph{color confinement}, the well-known difficulty in directly observing single color-charged entities, such as \emph{quarks}. Only when they combine to form more complex structures, like in \emph{mesons} and \emph{baryons}, can they be detected inside of space. Couldn't this be the signature of non-spatial states that cannot be drawn into space by a measurement, as if single quarks were the letters of a strange non-human language, and as if it was only when assembled together that they could form meaningful words, i.e. words that our spatial instruments would be able to understand, and therefore detect~\cite{Aerts2010}?

{\scshape Hudde:} This is a very suggestive image, on which I will have to meditate.

{\scshape Tielo:} My point is that if we adopt a view of reality which contemplates both discovery and creation processes, we cannot and should not decide in advance which are the possible spatial and non-spatial states, which among the non-spatial ones can be spatialized as opposed to those that cannot. On the other hand, if we assume, as Everettians usually do, that the Hilbert space is the only mathematical structure that should be used to represent the states of whatever physical entity, in whatever context, then we will conclude that all possible ``superpositions of worlds'' must exist in the multiverse. But you see, if you bring back into the theory the projection postulate, or rather, if you don't get rid of it in the first place and explain it by means of a hidden-measurement mechanism, you can no longer attribute such a unique role to the Hilbert space in the description of the states of physical entities. The more so, because if measurements characterized by non-uniform membranes are also taken into account, it is very easy to obtain probability models which can no longer be fitted into a Hilbertian structure~\cite{AertsSassoli2014a,AertsSassoli2014b}.

{\scshape Hudde:} Can you give me an example of superpositions that should exist according to a pure Hilbertian view, but that, from a more general, not necessarily Hilbertian view, might very well be inexistent? 

{\scshape Tielo:} We live surrounded by macroscopic objects of all kinds, which apparently do not show quantum effects. This chair on which I'm sitting, for example, is an entity in a well-defined spatial state, stably localized in this room. If the Hilbertian formalism is considered to be universal, then this same chair can in principle also exist, under the appropriate conditions, in a non-spatial state, obtained by superposing two states of the chair associated with two widely separated places. Can a chair exist in such a non-spatial, delocalized condition? According to the Everettian view, the answer, strictly speaking, is no, as the superposition state corresponds to two distinct chairs in two different locations, each one in a different parallel world. According to the standard interpretation of quantum mechanics, which considers that the linear Hilbert space is the only admissible state space, the answer is yes. Such delocalized state for the chair may be extremely difficult to produce, but in principle it is a possible state. Finally, according to the hidden-measurement interpretation, which doesn't attach to the Hilbert space any a priori fundamental role, both possibilities remain open, and the issue simply has to be decided on the basis of future experimental data.

{\scshape Hudde:} If I'm correct, superposition states of this non-spatial kind have already been obtained with entities like large organic molecules~\cite{Gerlich2011}. So don't you think that they should be possible also for macroscopic objects, although, of course, we may not be able to create the experimental conditions to produce them for a very long time, if at all?

{\scshape Tielo:} Well, to be sincere, I really don't know. But I'm convinced that to investigate fundamental questions of this kind, we had better not limit the structure of the state space from the beginning, and accept that more general mathematical structures can also play a role in our investigation of the entities populating our amazingly huge physical reality. 

{\scshape Hudde:} I see your point: this is apparently something we cannot do if we remain within the confines of a pure Hilbertian approach, be it with or without the projection postulate. 

{\scshape Tielo:} Exactly, we need approaches rich enough to account for the entire complexity of our reality. By the way, let me mention another feature of the many-worlds view that is problematic, at least from an intuitive point of view, and does not appear at all problematic in the many-measurements view. What about a chair that does not exist? 

{\scshape Hudde:} Are you asking what the many-worlds view says about entities that do not exist? 

{\scshape Tielo:} Precisely. We talked about the possibility for a chair to be in a non-spatial state. What about a non-existing chair? Don't you think that in our construction of reality we should be able to also distinguish existence from non-existence? That is, chairs that are present in this room and chairs that are not present in it, not because they are in some non-spatial state, or because they are located somewhere else in space, but simply because they have not yet been built and might possibly never be built, hence only exist as a possibility.

{\scshape Hudde:} I understand, in the many-worlds interpretation all the things that can exist, in accordance with the laws of physics, also exist in at least one of the universes. Hence, there is no such thing as a \emph{possible chair}, because in one of the universes it will necessarily exist, if the laws of physics allow it to exist. In the many-worlds interpretation it is indeed quite difficult, if not impossible, to distinguish between existence and non-existence.

{\scshape Tielo:} I must admit that when I became aware of the fact that the many-measurements approach can handle the notion of \emph{existence} in a non problematic way, allowing it to be different from the notion of \emph{possible existence}, in the sense of not having to think about possible existence as existence in another universe, this was for me one of its obvious points of attractiveness as compared to the many-worlds interpretation. Clearly, this lack of distinction between existence and possible existence in the many-worlds interpretation is due to the choice of describing superposition states as collections of collapsed states in the different universes. Think again of the table-worlds example. In the many-worlds interpretation, an off-table die is represented by a collection of different dice, placed on different tables, in all the possible on-table states that one can obtain by throwing the die on a table. On the other hand, as we observed already, in the many-measurements view an off-table die is also allowed to exist, and corresponds to a real superposition state; but also, if such off-table die is thrown on a table, it will only exist as an on-table die relative to that specific table: in no way will it exist as an on-table die relative to the other tables on which it could have been rolled. In this sense, although the many-measurements view, similarly to the many-worlds view, considers reality to be much bigger than its collapsed regions, there is no internal logic forcing it to introduce this intuitively problematic relation with non-existence. Dice can be in genuine superposition states, when not thrown on any table, but we can also think of dice that do not exist at all, although the physical laws would not forbid their existence: dice that are not in superposition states, or in collapsed states, are just `possible dice' and nothing more. Similarly for chairs, we can choose to have an experience with one of the chairs present in this room, or in some other places, by choosing to sit on it, and in principle we can also have a ``sitting experience'' with a hypothetical chair in some non-spatial state, although in this case the outcome of the experience will not be certain in advance; but with certainty we will never be able to sit on a non-existing chair, that is, on a chair that has not yet been created by a carpenter and that perhaps will never be created. 

{\scshape Hudde:} Yes, whereas in the many-worlds reality we always have the possibility of ``sitting on a non-existing chair,'' in the sense that all the chairs that can in principle be created have been created, somewhere in the multiverse. 

{\scshape Tielo:} That's right, and also, all the chairs that can in principle be destroyed have not been destroyed, somewhere in the multiverse. By the way, let me say that this issue of distinguishing between existing and non-existing entities appears not to be a purely academic one. For instance, when we address the difficult problem of finding a common consistent framework for quantum theory and relativity, a clear distinction between created and uncreated elements of our reality seems to be of fundamental importance, particularly on the issue of determining whether part of our future reality also exists in our present, as I got to learn by reading some of the works I mentioned to you~\cite{Aerts1999}. But it would take a little too long to explain  this to you, so let us leave this for a possible future discussion.
 
{\scshape Hudde:} Agreed. Well, let me say again that this has been a very engaging conversation. From a many-worlds perspective, I'm certain I will bifurcate soon into a world where we will meet again and continue our exchange of ideas. And from a many-measurements perspective, I just hope we will soon create this opportunity. 

{\scshape Tielo:} Well said, and see you soon anyway.
\newline

A final remark is in order. The main purpose of this dialogue was to present some of the ideas and results of the hidden-measurement interpretation of quantum mechanics by mildly opposing them to those usually defended by many-worlders. We do not pretend to have presented the latter in a comprehensive way or to have taken their defense with a lot of conviction; this is not only because it was not our purpose to do so, but also because different conflicting interpretations seem to co-exist regarding how one should understand and implement Everett's program~\cite{Everett1957,Deutsch1999,Deutsch1996, DeWitt1973,Hartle1968,Geroch1984,Albert1988,Lockwood1996,Gell-Mann1993,Saunders2010}. We nevertheless hope that Everettians and, more generally, physicists and philosophers interested in foundational issues, will be stimulated by the above dialogue and decide -- as one of the characters apparently did -- to take a closer look at the hidden-measurement (one-world) approach and the solution it offers to the measurement problem~\cite{AertsSassoli2014c}. We also look forward to receiving any comments to the present exchange, so that the conversation can continue, be it in this or another world.

\end{document}